\documentclass[superscriptaddress, twocolumn]{revtex4}
\usepackage{amsmath}
\usepackage{graphicx}

\usepackage{floatflt,epsfig}
\usepackage{rotating}
\begin{document}

\def\ii{\'{\i}}
\def\d{\mbox{d}}

\title{Electrically charged pulsars}

\author{M.D. Alloy}
\affiliation{ Depto de F\ii sica - CFM - Universidade Federal de Santa Catarina  Florian\'opolis - SC - CP. 476 - CEP 88.040 - 900 - Brazil}
\author{D.P. Menezes}
\affiliation{ Depto de F\ii sica - CFM - Universidade Federal de Santa Catarina  Florian\'opolis - SC - CP. 476 - CEP 88.040 - 900 - Brazil}

\begin{abstract}
In the present work we investigate one possible variation on the usual
electrically neutral pulsars: the inclusion of electric charge. We study the effect of 
electric charge in pulsars assuming that the charge distribution is 
proportional to the energy density. All calculations were performed for zero
temperature and fixed entropy equations of state. 
\end{abstract}

\maketitle

\vspace{0.5cm}
PACS number(s):26.60.+c,24.10.Jv,21.65.+f,95.30.Tg 
\vspace{0.5cm}

\section{Introduction}

Pulsars are believed to be the remnants of supernova explosions. They have 
masses 1--$2M_\odot$, radii $\sim10\rm\,km$, and a temperature of the 
order of $10^{11}\rm\,K$ at birth, cooling within a few days to about 
$10^{10}\rm\,K$ by emitting neutrinos. Pulsars are normally known as neutron 
stars. Qualitatively, a neutron star is analogous to a white dwarf star, 
with the pressure due to degenerate neutrons rather than degenerate 
electrons. The assumption that the 
neutrons in a neutron star can be treated as an ideal gas is not well 
justified: the effect of the strong force needs to be taken into 
account by replacing the equation of state (EoS) for an ideal gas by 
a more realistic EoS. The composition of pulsars  
remains a source of speculation, with some of the possibilities being 
the presence of hyperons \cite{aquino,magno,rafael}, a mixed phase 
of hyperons and quarks \cite{mp1,mp,pmp1,pmp2,trapping}, a phase of 
deconfined quarks or pion and kaon condensates \cite{kaons}. Another
possibility would be that pulsars are, in fact, quark stars \cite{quark}.
In conventional models, hadrons are assumed to be the true ground state of the 
strong interaction. However, it  has been argued 
\cite{itoh,bodmer,witten,haensel,olinto} that {\it strange matter}
composed of deconfined $u,d$ and $s$ quarks is the true ground  state of all 
matter. In the stellar modeling, the structure of the star depends on the 
assumed EoS, which is different in each of the above mentioned cases. 
An important distinction 
between quark stars and conventional neutron stars is that the quark 
stars are self-bound by the strong interaction, whereas neutron stars 
are bound by gravity. 

Once an adequate EoS is chosen, it is used as input to the 
Tolman-Oppenheimer-Volkoff (TOV) equations \cite{tov}, which are derived from 
Einstein's equations in the Schwarzschild metric for a static, 
spherical star. Some of the stellar properties, as the radius, gravitational
and baryonic masses, central energy densities, etc are obtained. These
results are then tested against some of the constraints provided by
astronomers and astrophysicists \cite{cottam,sanwal} and some of the EoS are
shown to be inappropriate for describing pulsars \cite{mp,pmp1,kaons}.

Notice also that the temperature in the interior of the star is 
not constant \cite{burrows,trapping}, but the entropy per baryon is. This is 
the reason for choosing fixed entropies to take the temperature effects into 
account. The maximum entropy  per baryon ($S$) reached in the core of 
a new born star is about 2 (in units of Boltzmann's constant) \cite{prak97}.
We then use EoS obtained with $S=0~(T=0$), 1 and 2.

We study the effects of the electric charge in compact stars. 
This study was first performed in stars composed of hot ionized gas \cite{idea} 
and then reconsidered for a cold star ($T=0$) described by a polytropic 
EoS \cite{mane}. The electric charge distribution is assumed to be 
proportional to the mass density. The TOV equations again have to 
be modified to take the electric field into account. In the present 
work once more all possible classes of pulsars are examined 
in the presence of the electric field for $S=0,1$ and $2$.

This paper is organized as follows: in Sec. II the formalisms of the
electrically charged stars are revisited and the results are presented. 
In Sec. III the results are discussed and the main conclusions are drawn.

\section{Formalism and Results}

As the first step we need to know the EoS of the system,
\[
\epsilon=\epsilon(p),\qquad n=n(p),
\]
where $p$ is the pressure, $\epsilon$ is the energy density, and $n$ is the 
number density of baryons. Once an 
adequate EoS is obtained, it can be used to provide the stellar properties.

\subsection{Electrically charged compact stars}

In this section we include modifications in the TOV equations to describe 
electrically charged pulsars with null angular velocity.
The geometry that describes a static spherical star is given by equation (1). In order that the Maxwell equations are incorporated into the stress tensor $T_\nu^\mu$, it becomes:
\begin{equation}
T_\nu^\mu=(p+\epsilon)u^\mu u_\nu-p\delta^\mu_\nu+\frac{1}{4\pi}\left(F^{\mu\alpha}F_{\alpha\nu}-\frac{1}{4}\delta^\mu_\nu F_{\alpha\beta}F^{\alpha\beta}\right),
\end{equation}
where again $p$ is the pressure, $\epsilon$ is the energy density, and 
$u^\mu$ is the 4-velocity vector.

The electromagnetic field obeys the relation
\begin{equation}
  \left[\sqrt{-g}F^{\mu \nu}\right]_{,\nu}=4\pi j^{\mu}\sqrt{-g},
\end{equation}
where $j^\mu$ is the four current density. Next we consider static 
stars only. Hence the electromagnetic field is only  due to the electric 
charge, which means that $F^{01}=-F^{10}$, and the other terms are absent. 
From the four-potential $A_{\mu}$ the surviving potential is $A_0=\phi$.
Thus the electric field is given by
\begin{equation}
E(r)=\frac{1}{r^2}\int^r_04\pi r^2 j^0e^{(\nu+\lambda)/2}dr,
\end{equation}
where $j^0e^{\nu/2}=\rho_{ch}$ is the charge density. The electric field can 
be written as
\begin{equation}
\frac{dE(r)}{dr}=-\frac{2E}{r}+4\pi\rho_{ch}e^{\lambda/2}
\end{equation}
and total charge of the system as
\begin{equation}
Q=\int^R_04\pi r^2\rho_{ch}e^{\lambda/2}dr,
\end{equation}
where $R$ is the radius of the star.

In the star frame the mass is
\begin{equation}
\frac{dM_{tot}}{dr}(r)=4\pi r^2\left(\epsilon+\frac{E(r)^2}{8\pi}\right).
\end{equation}

To an observer at infinity, the mass is
\begin{equation}
M_\infty=\int^{\infty}_04\pi r^2\left(\epsilon+\frac{E(r)^2}{8\pi}\right)dr=M_{tot}(R)+\frac{Q(R)^2}{2R}
\end{equation}

By using the conservation law of the stress tensor $(T^\mu_{\nu;\mu}=0)$ we obtain the hydrostatic equation
\begin{eqnarray}
\frac{dp}{dr}&=&-\frac{\left[M_{tot}+4\pi r^3\left(p-\frac{E(r)^2}{8\pi}\right)\right](\epsilon+p)}{r^2\left(1-\frac{2M_{tot}}{r}\right)}\nonumber\\
&+&\rho_{ch}E(r)e^{\lambda/2}.
\end{eqnarray}
The first term on the right-hand side comes from the gravitational force and the second term comes from the Coulomb force.
By using the metric and the relation
\begin{equation}
R^\mu_\nu-\frac{1}{2}R\delta^\mu_\nu=-8\pi T^\mu_\nu,
\end{equation}
we obtain the following differential equation
\begin{equation}
\frac{d\lambda}{dr}=\left[8\pi re^{\lambda}\left(\epsilon+\frac{E(r)^2}{8\pi}\right)-\left(\frac{e^\lambda-1}{r}\right)\right],
\end{equation}
which is used to determine the metric $e^\lambda$.

\begin{figure*}
\includegraphics[width=4.0in, height=2.3in, angle=270]{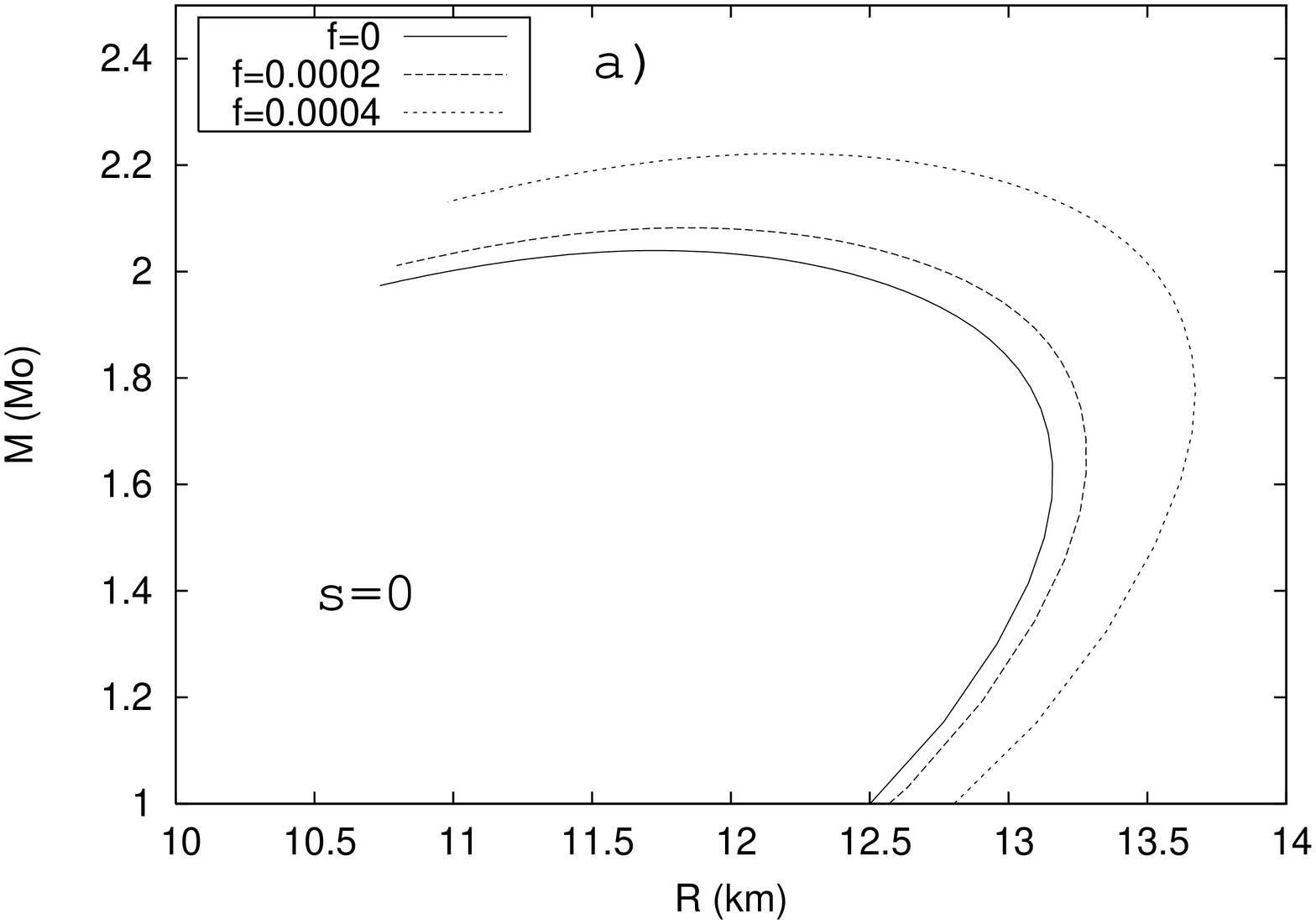}
\includegraphics[width=4.0in, height=2.3in, angle=270]{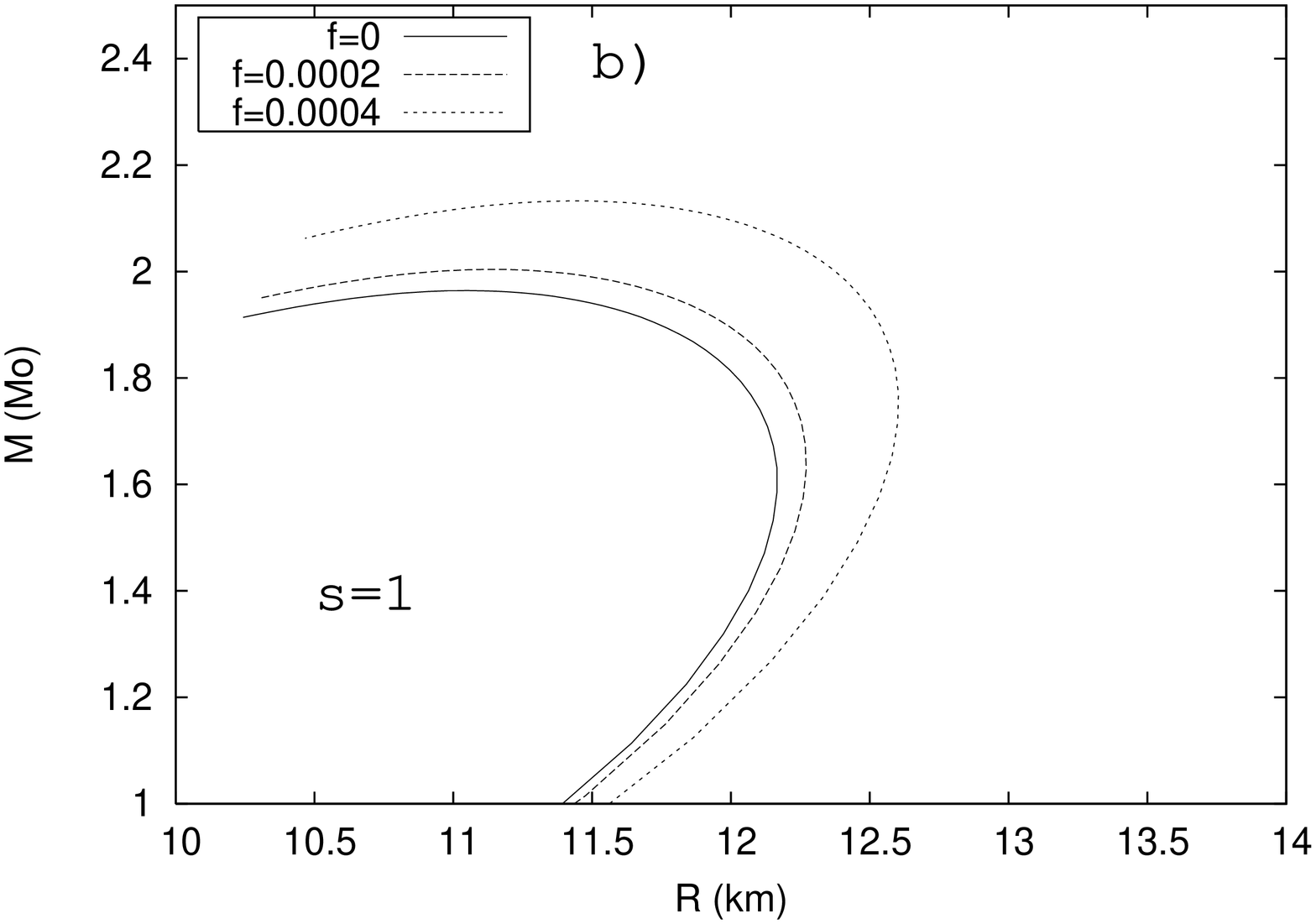}
\includegraphics[width=4.0in, height=2.3in, angle=270]{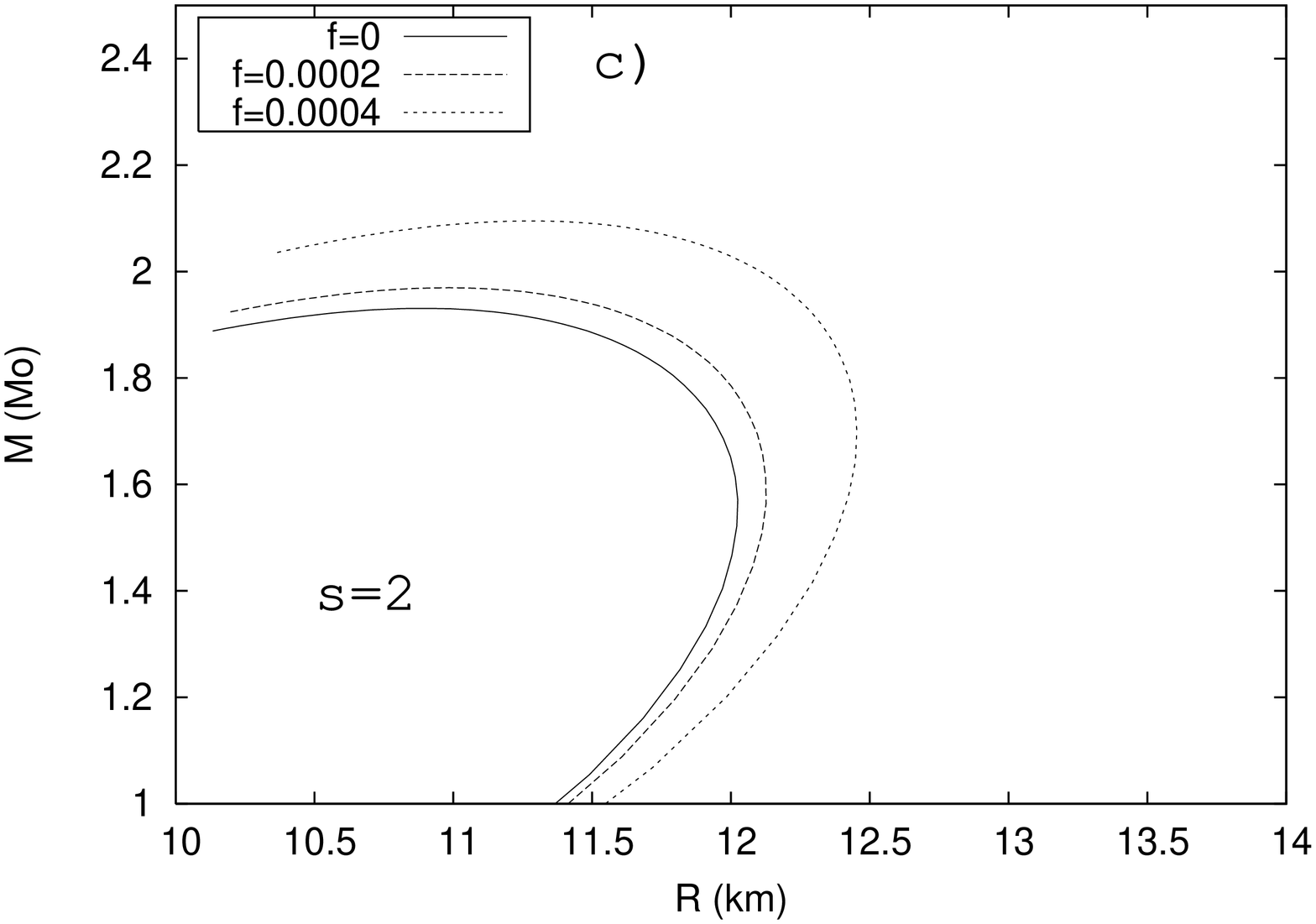}
\caption{Solutions for electrically charged hadronic stars with different 
values of $f$.}
\label{fig1}
\end{figure*}

\begin{figure*}
\includegraphics[width=4.0in, height=2.3in, angle=270]{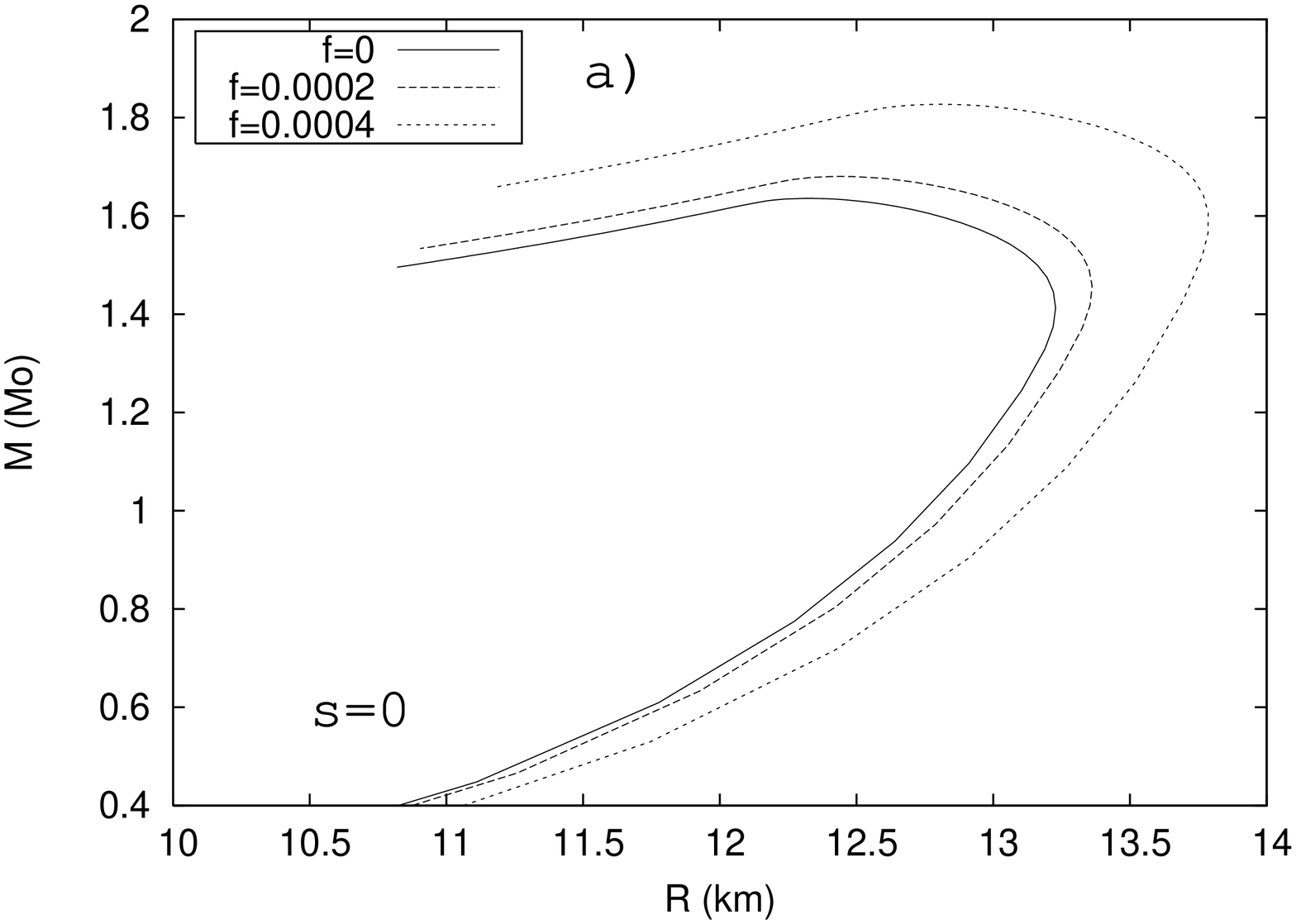}
\includegraphics[width=4.0in, height=2.3in, angle=270]{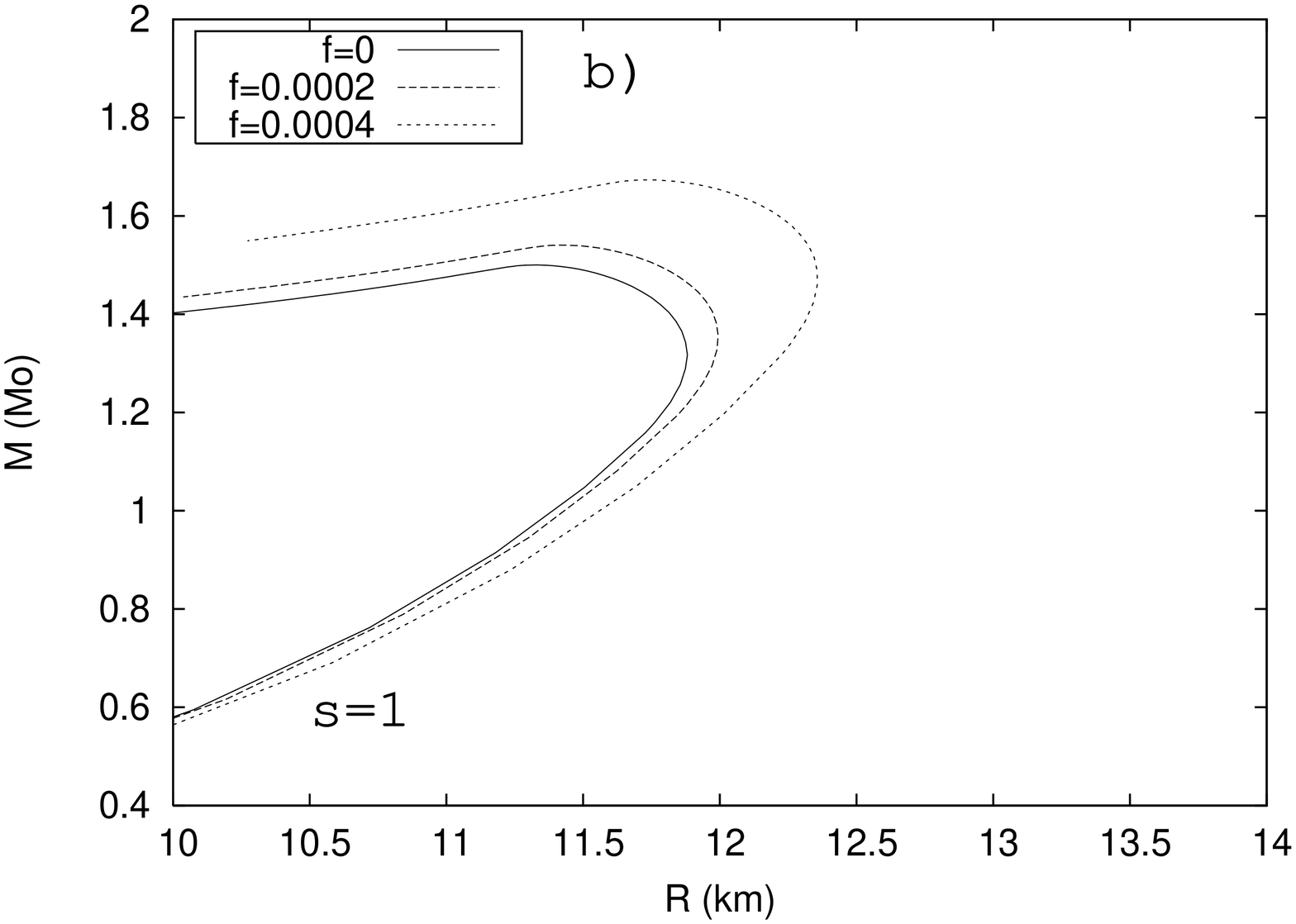}
\includegraphics[width=4.0in, height=2.3in, angle=270]{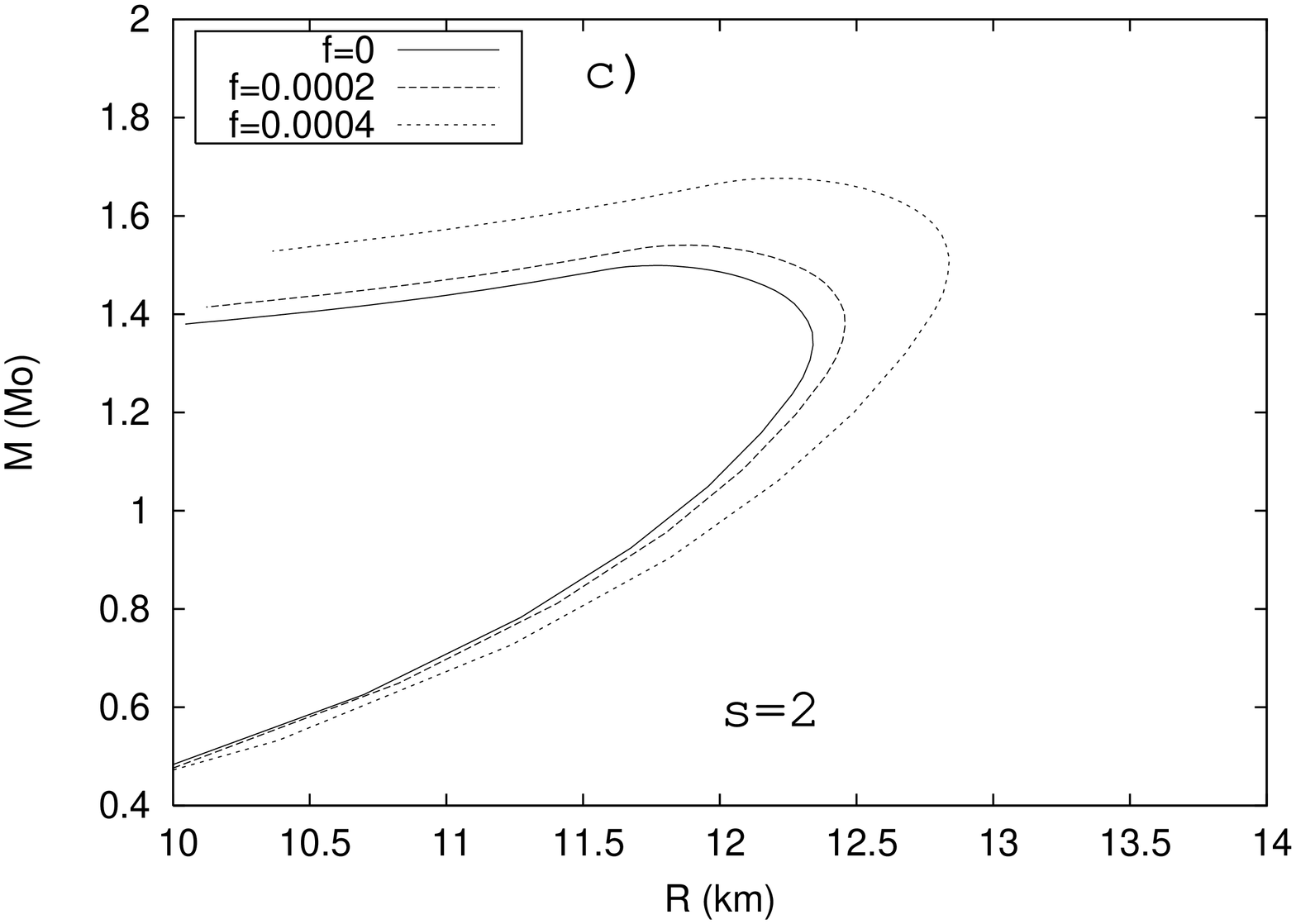}
\caption{Solutions for electrically charged hybrid stars with
  different values of $f$.}
\label{fig2}
\end{figure*}

\begin{figure*}
\includegraphics[width=4.0in, height=2.3in, angle=270]{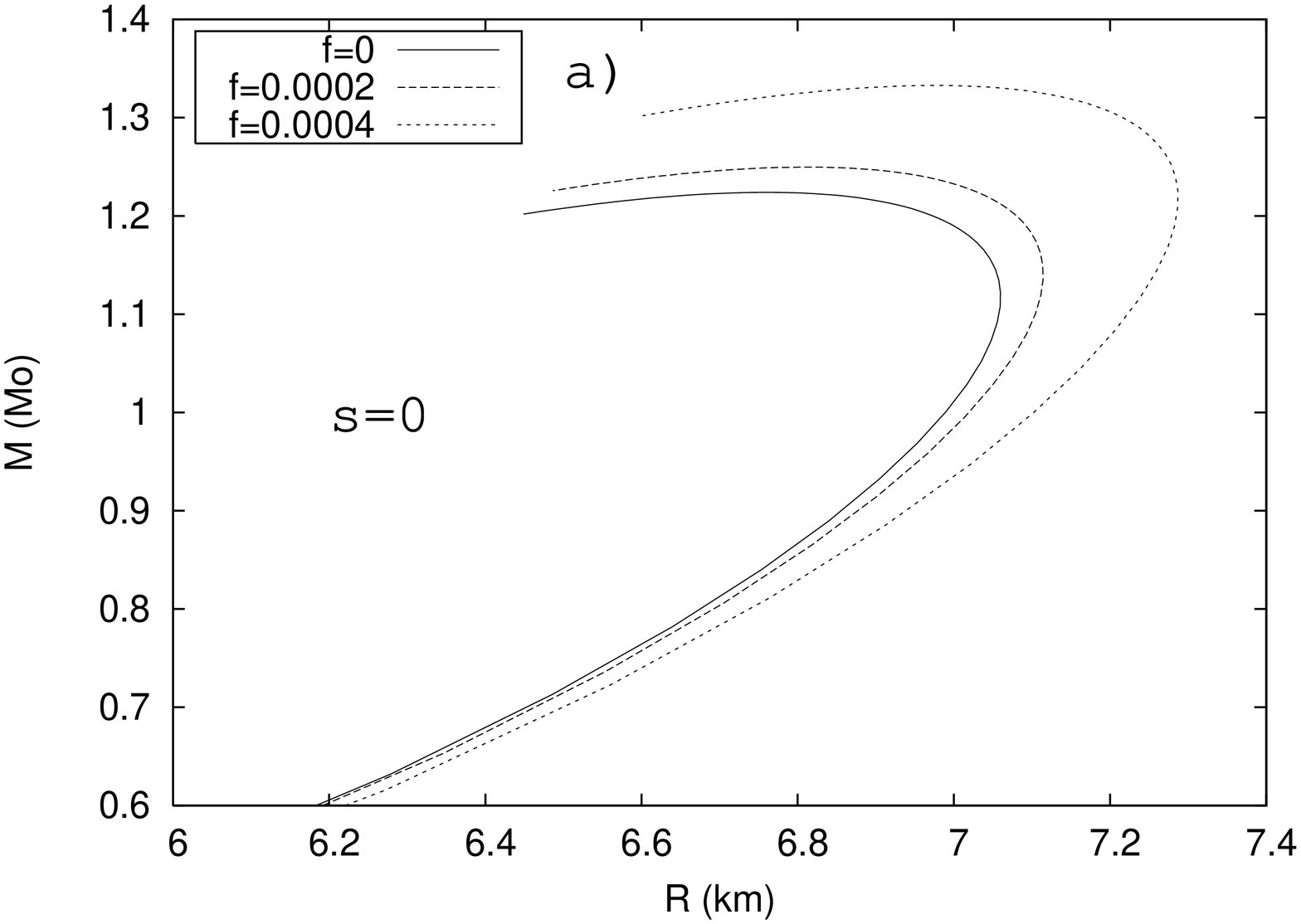}
\includegraphics[width=4.0in, height=2.3in, angle=270]{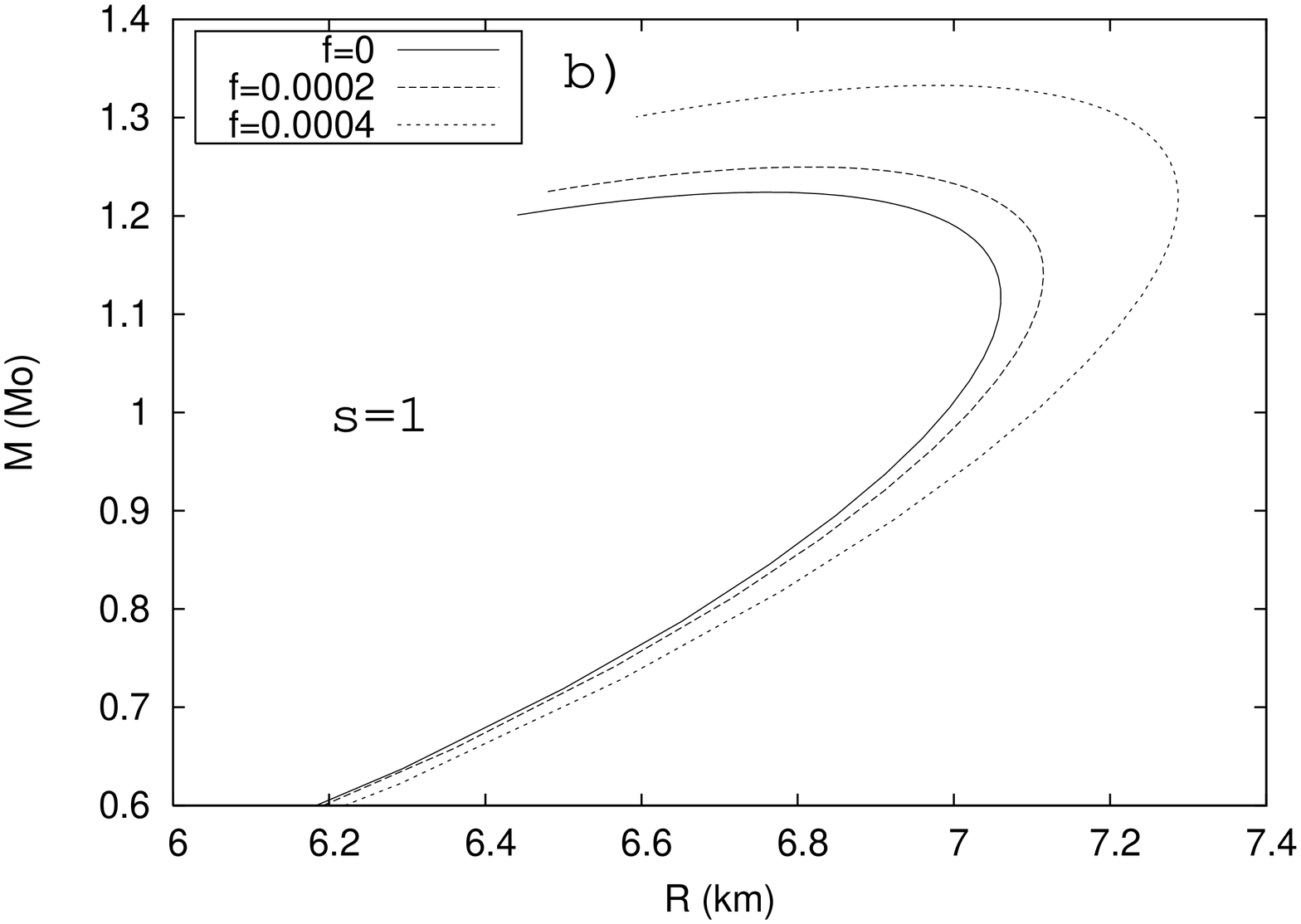}
\includegraphics[width=4.0in, height=2.3in, angle=270]{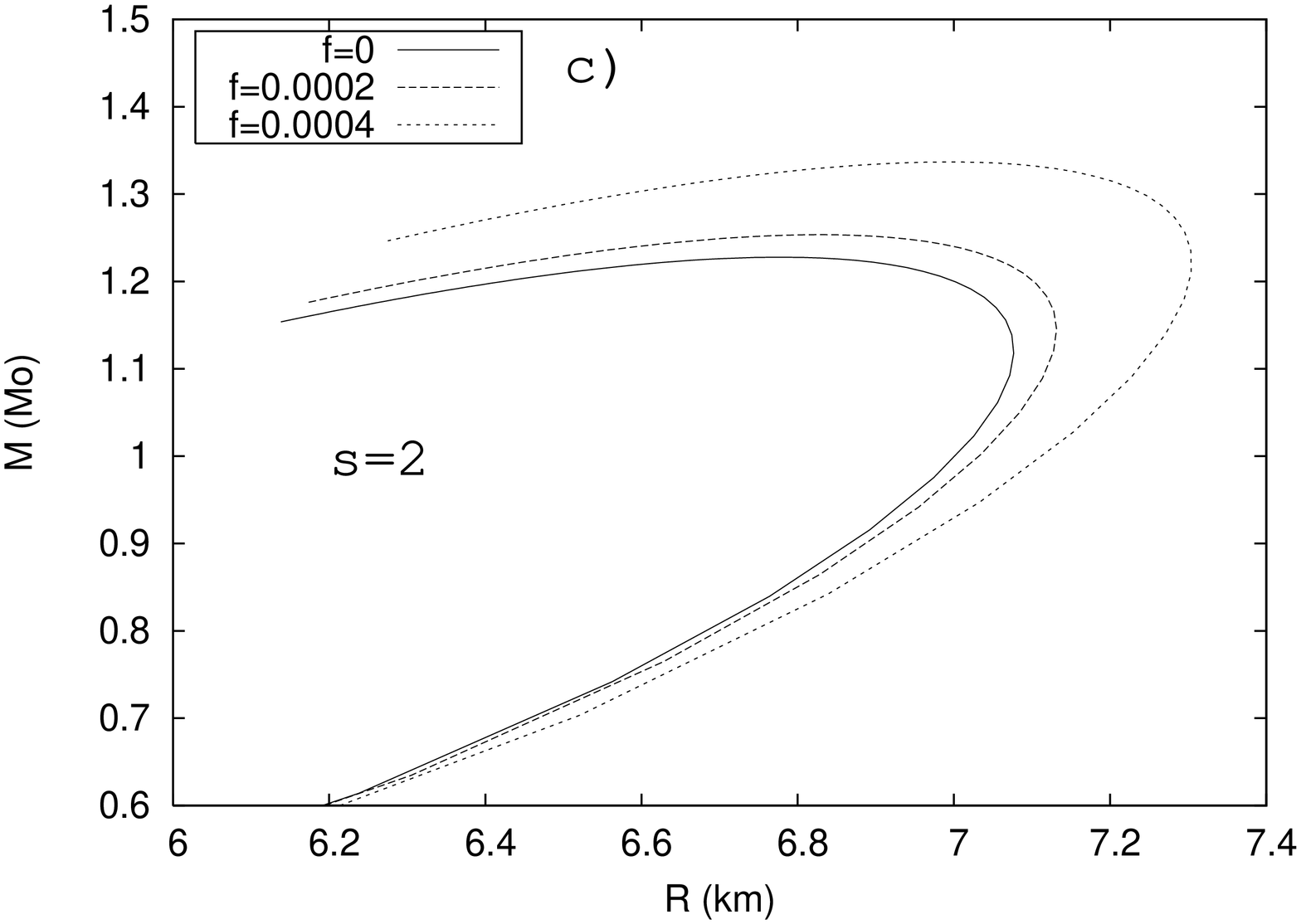}
\caption{Solutions for electrically charged quarkionic stars obtained with the 
MIT bag model for different values of $f$.}
\label{fig3}
\end{figure*}

\begin{figure*}
\includegraphics[width=4.0in, height=2.3in, angle=270]{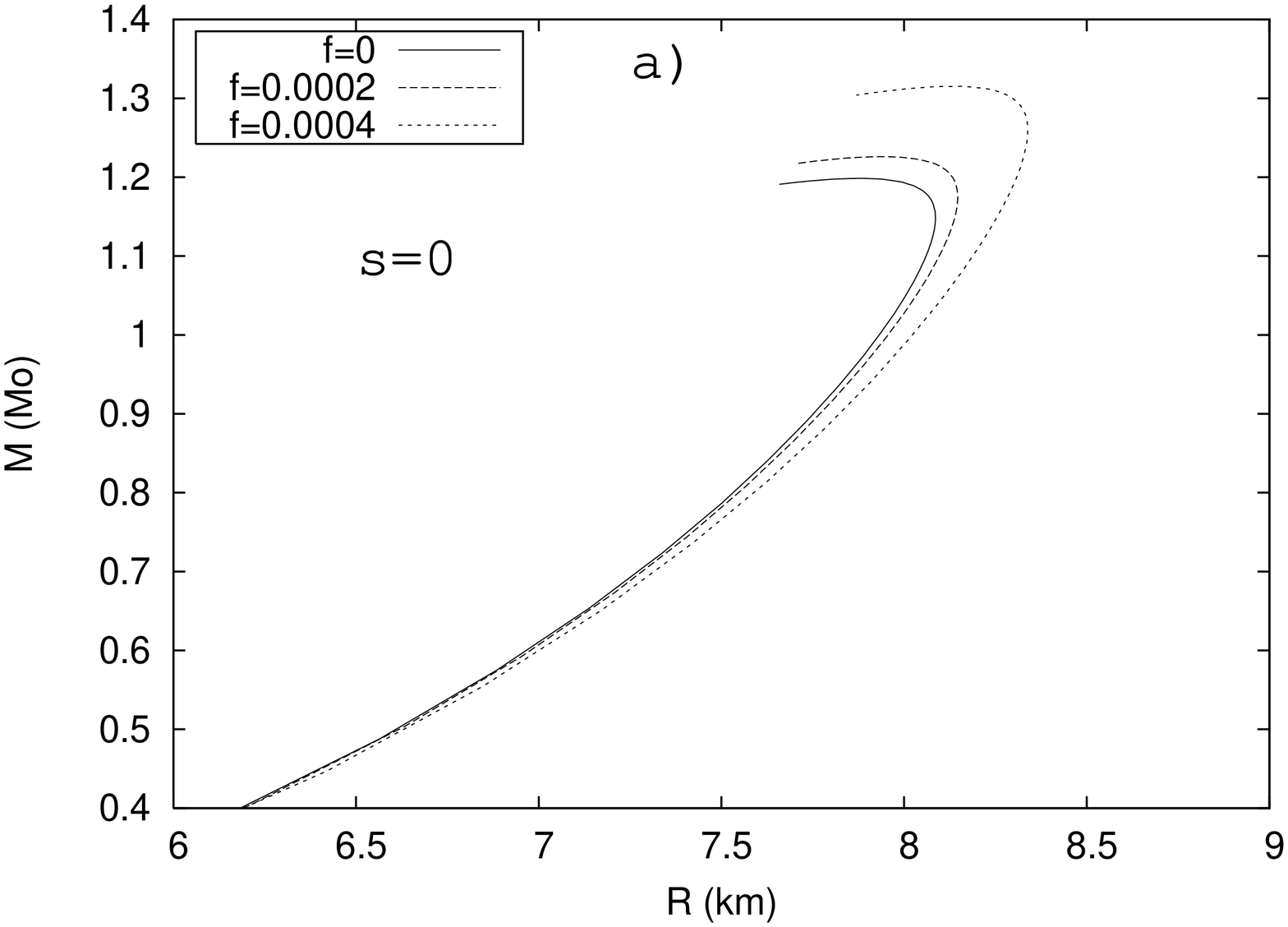}
\includegraphics[width=4.0in, height=2.3in, angle=270]{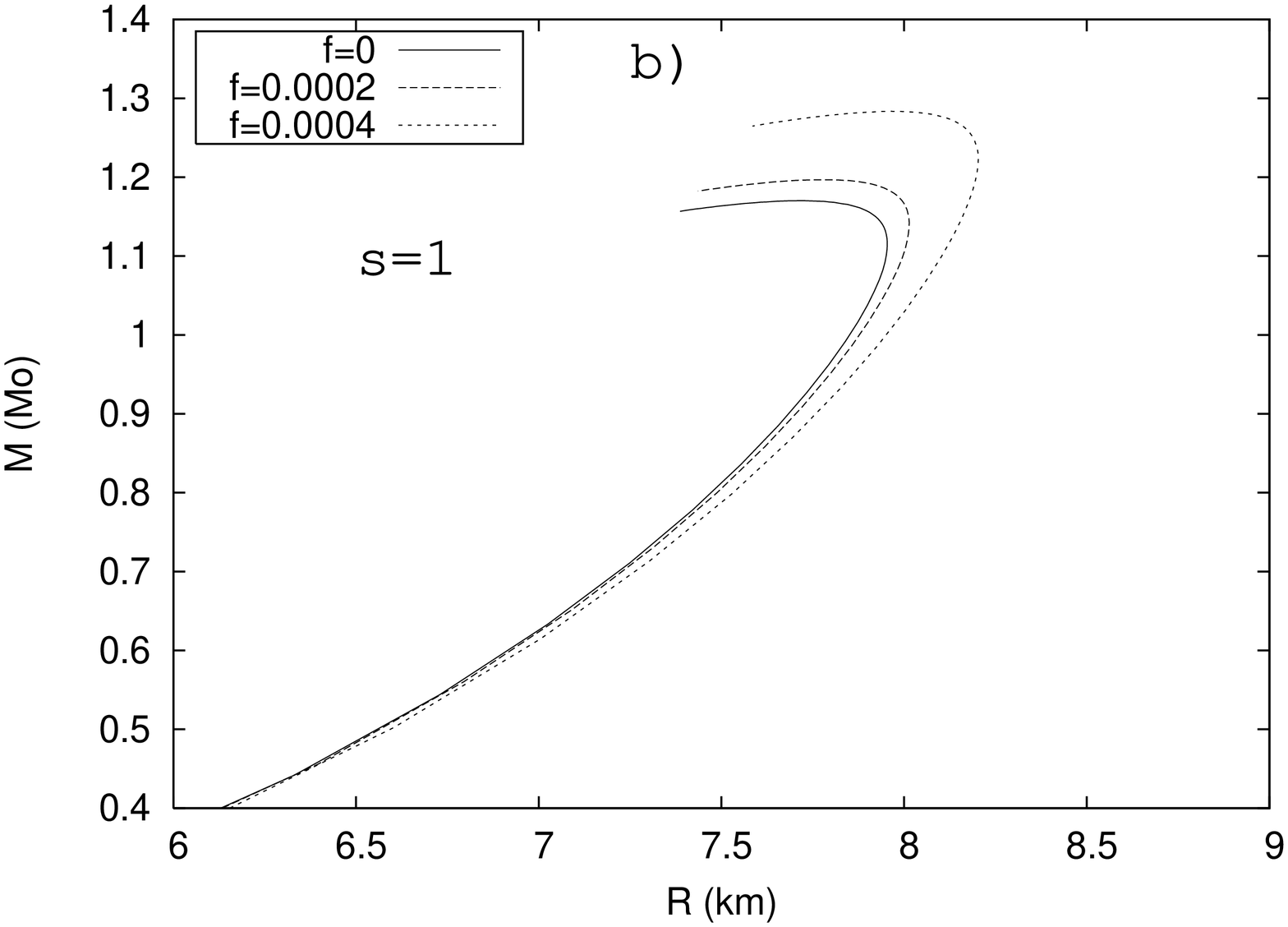}
\includegraphics[width=4.0in, height=2.3in, angle=270]{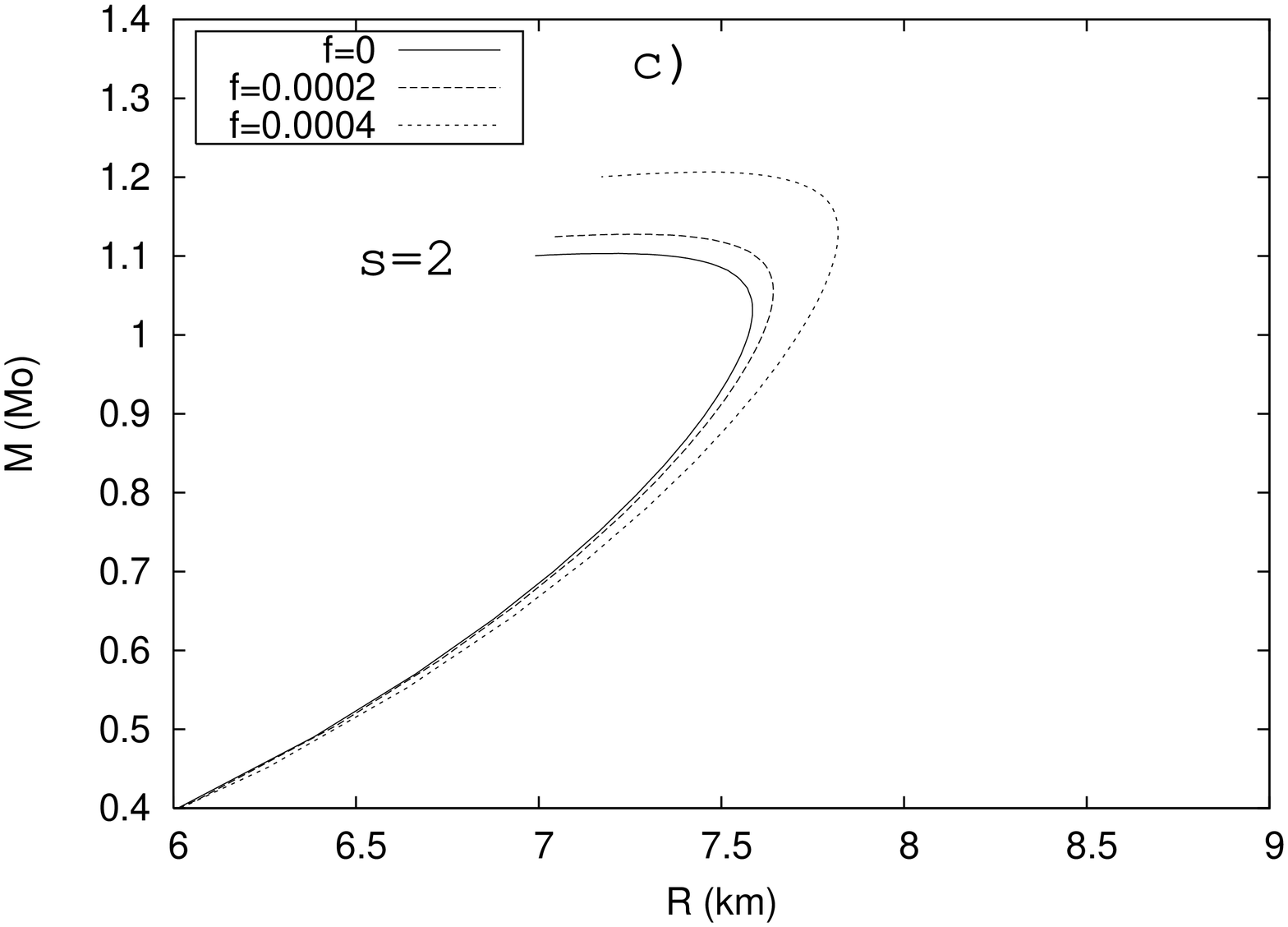}
\caption{Solutions for electrically charged quarkionic stars obtained with the 
NJL model for different values of $f$.}
\label{fig4}
\end{figure*}

So, we have a set of differential equations to be solved formed by 
equations(19), (21), (23) and (25).
The boundary conditions at $r=0$ are $E(r)=0$, $e^\lambda=1$,
$n=\rho_c$ and at $r=R$, $p=0$. We assume that the charge goes with the 
energy density $\epsilon$ as prescribed in \cite{mane}:

\begin{equation}
\rho_{ch}=f\times0.86924\times10^{3}\epsilon.
\end{equation}
This choice of charge distribution is a reasonable assumption in the sense 
that a large mass can hold a large amount of charge.
In table \ref{tab:tab2} results for electrically charged neutron stars are 
presented. We have calculated the results for 48 different configuration 
models of compact stars. 
Once again the EoS for hadronic and hybrid stars were taken from 
\cite{trapping}, the EoS for quarkionic stars were taken from \cite{quark}.
In table \ref{tab:tab2} the electric charge $Q$ is given in Coulomb 
and $f$ varies from zero (no charge) to a small value (0.0006).
The related mass-radius plots for hadronic, hybrid and quarkionic stars are 
given respectively in figs. \ref{fig1},\ref{fig2} and \ref{fig3}-\ref{fig4}.

\begin{table*}[h]
\begin{center}
\caption{\label{tab:tab2}Electrically compact stars with different charge fraction $f$.}
\begin{ruledtabular}
\begin{tabular}{ccccccccc}
\hline
Type & Entropy & $f$ & $M_{\max}$ & $M_{\infty}$ & $R$ & $\epsilon_c$ & $Q$\\
&&& $(M_{\odot})$ & $(M_{\odot})$ & $(km)$  & $(g/cm^3)$ & $(C)$\\
\hline
Hadronic & 0 & 0 & 2.04 & 2.04 & 11.72 & $1.98\times 10^{15}$ & $0$ \\
Hadronic & 0 & 0.0002 & 2.08 & 2.10 & 11.84 & $1.94\times 10^{15}$ & $7.97\times 10^{19}$ \\
Hadronic & 0 & 0.0004 & 2.22  & 2.28 & 12.18 & $1.84\times 10^{15}$ & $1.71\times 10^{20}$\\
Hadronic & 0 & 0.0006 & 2.50  & 2.66 & 12.73 & $1.75\times 10^{15}$ & $2.91\times 10^{20}$\\
Hadronic & 1 & 0 & 1.96 & 1.96 & 11.02 & $2.23\times 10^{15}$ & $0$ \\
Hadronic & 1 & 0.0002 & 2.00 & 2.02 & 11.15 & $2.13\times 10^{15}$ & $7.69\times 10^{19}$ \\
Hadronic & 1 & 0.0004 & 2.13  & 2.19 & 11.44 & $2.04\times 10^{15}$ & $1.64\times 10^{20}$\\
Hadronic & 1 & 0.0006 & 2.39  & 2.55 & 12.01 & $1.85\times 10^{15}$ & $2.78\times 10^{20}$\\
Hadronic & 2 & 0 & 1.93 & 1.93 & 10.91 & $2.24\times 10^{15}$ & $0$ \\
Hadronic & 2 & 0.0002 & 1.97 & 1.98 & 11.01 & $2.19\times 10^{15}$ & $7.55\times 10^{19}$ \\
Hadronic & 2 & 0.0004 & 2.09  & 2.15 & 11.26 & $2.15\times 10^{15}$ & $1.61\times 10^{20}$\\
Hadronic & 2 & 0.0006 & 2.34  & 2.50 & 11.86 & $1.90\times 10^{15}$ & $2.72\times 10^{20}$\\
\hline
hybrid & 0 & 0 & 1.64  & 1.64 & 12.33 & $1.57\times 10^{15}$ & $0$\\
hybrid & 0 & 0.0002 & 1.68  & 1.69 & 12.43 & $1.57\times 10^{15}$ & $5.98\times 10^{19}$\\
hybrid & 0 & 0.0004 & 1.82  & 1.86 & 12.82 & $1.48\times 10^{15}$ & $1.31\times 10^{20}$\\
hybrid & 0 & 0.0006 & 2.13  & 2.23 & 13.52 & $1.39\times 10^{15}$ & $2.31\times 10^{20}$\\
hybrid & 1 & 0 & 1.50  & 1.50 & 11.32 & $1.75\times 10^{15}$ & $0$\\
hybrid & 1 & 0.0002 & 1.54  & 1.55 & 11.43 & $1.71\times 10^{15}$ & $5.44\times 10^{19}$\\
hybrid & 1 & 0.0004 & 1.67  & 1.70 & 11.74 & $1.66\times 10^{15}$ & $1.19\times 10^{20}$\\
hybrid & 1 & 0.0006 & 1.94  & 2.03 & 12.34 & $1.58\times 10^{14}$ & $2.10\times 10^{20}$\\
hybrid & 2 & 0 & 1.50 & 1.50 & 11.76 & $1.58\times 10^{15}$ & $0$\\
hybrid & 2 & 0.0002 & 1.54  & 1.55 & 11.86 & $1.58\times 10^{15}$ & $5.41\times 10^{19}$\\
hybrid & 2 & 0.0004 & 1.68  & 1.71 & 12.21 & $1.53\times 10^{15}$ & $1.18\times 10^{20}$\\
hybrid & 2 & 0.0006 & 1.95  & 2.04 & 12.86 & $1.44\times 10^{14}$ & $2.10\times 10^{20}$\\
\hline
Quarkonic(MIT) & 0 & 0 & 1.22  & 1.22 & 6.77 & $5.14\times 10^{15}$ & $0$\\
Quarkonic(MIT) & 0 & 0.0002 & 1.25  & 1.26 & 6.81 & $5.13\times 10^{15}$ & $4.75\times 10^{19}$\\
Quarkonic(MIT) & 0 & 0.0004 & 1.33  & 1.37 & 6.97 & $4.95\times 10^{15}$ & $1.02\times 10^{20}$\\
Quarkonic(MIT) & 0 & 0.0006 & 1.50  & 1.60 & 7.28 & $4.56\times 10^{15}$ & $1.74\times 10^{20}$\\
Quarkonic(MIT) & 1 & 0 & 1.22  & 1.22 & 6.76 & $5.17\times 10^{15}$ & $0$\\
Quarkonic(MIT) & 1 & 0.0002 & 1.25  & 1.26 & 6.82 & $5.07\times 10^{15}$ & $4.75\times 10^{19}$\\
Quarkonic(MIT) & 1 & 0.0004 & 1.33  & 1.37 & 6.98 & $4.88\times 10^{15}$ & $1.02\times 10^{20}$\\
Quarkonic(MIT) & 1 & 0.0006 & 1.50  & 1.60 & 7.29 & $4.50\times 10^{15}$ & $1.74\times 10^{20}$\\
Quarkonic(MIT) & 2 & 0 & 1.23  & 1.23 & 6.79 & $5.08\times 10^{15}$ & $0$\\
Quarkonic(MIT) & 2 & 0.0002 & 1.25  & 1.26 & 6.83 & $5.09\times 10^{15}$ & $4.77\times 10^{19}$\\
Quarkonic(MIT) & 2 & 0.0004 & 1.34  & 1.37 & 6.98 & $4.94\times 10^{15}$ & $1.03\times 10^{20}$\\
Quarkonic(MIT) & 2 & 0.0006 & 1.50  & 1.60 & 7.31 & $4.49\times 10^{15}$ & $1.74\times 10^{20}$\\
\hline
Quarkonic(NJL) & 0 & 0 & 1.20  & 1.20 & 7.87 & $3.45\times 10^{15}$ & $0$\\
Quarkonic(NJL) & 0 & 0.0002 & 1.23  & 1.23 & 7.93 & $3.45\times 10^{15}$ & $4.42\times 10^{19}$\\
Quarkonic(NJL) & 0 & 0.0004 & 1.32  & 1.34 & 8.14 & $3.30\times 10^{15}$ & $9.54\times 10^{19}$\\
Quarkonic(NJL) & 0 & 0.0006 & 1.49  & 1.57 & 8.48 & $3.20\times 10^{15}$ & $1.64\times 10^{20}$\\
Quarkonic(NJL) & 1 & 0 & 1.17  & 1.17 & 7.71 & $3.68\times 10^{15}$ & $0$\\
Quarkonic(NJL) & 1 & 0.0002 & 1.19  & 1.20 & 7.79 & $3.56\times 10^{15}$ & $4.32\times 10^{19}$\\
Quarkonic(NJL) & 1 & 0.0004 & 1.28  & 1.31 & 8.00 & $3.50\times 10^{15}$ & $9.31\times 10^{19}$\\
Quarkonic(NJL) & 1 & 0.0006 & 1.45  & 1.53 & 8.31 & $3.38\times 10^{15}$ & $1.60\times 10^{20}$\\
Quarkonic(NJL) & 2 & 0 & 1.10  & 1.10 & 7.18 & $4.58\times 10^{15}$ & $0$\\
Quarkonic(NJL) & 2 & 0.0002 & 1.13  & 1.13 & 7.29 & $4.34\times 10^{15}$ & $4.09\times 10^{19}$\\
Quarkonic(NJL) & 2 & 0.0004 & 1.20  & 1.23 & 7.42 & $4.40\times 10^{15}$ & $8.23\times 10^{19}$\\
Quarkonic(NJL) & 2 & 0.0006 & 1.36  & 1.43 & 7.80 & $3.98\times 10^{15}$ & $1.50\times 10^{20}$\\
\hline
\end{tabular}
\end{ruledtabular}
\end{center}
\end{table*}

\section{Results and Conclusions}

Let's now go back to our results in order to compare them with what is found
in the literature and draw the conclusions. 

In \cite{antigo1} the
models named O,P,Q and R given in table \ref{tab:tab2} 
can again be compared with our result for the hadronic star at $T=0$ 
and they are indeed very similar. 

We next look at the results displayed in table \ref{tab:tab2} and corresponding
figures. The general trend is the same observed in a simple polytropic EoS
for $T=0$ \cite{mane}, i.e., the electric charge, the maximum mass and
the mass observed at infinity increase with $f$, as it should be. Although the
EoS used in the present work are very different from the one used in
\cite{mane} the values of the radii obtained and the electric charge for a 
fixed $f$ value are compatible. Figs. \ref{fig1},\ref{fig2},\ref{fig3} and
\ref{fig4} also show the same behavior as fig. 2 of \cite{mane}, i.e., as $f$
increases, the maximum mass and radius of a family of stars increase.
The EoS used in the present work are for bare pulsars, i.e. the outmost layer
is not included. From table \ref{tab:tab2} one can see that 
the effect of entropy on a charged star
remains the same as in a neutral star: the maximum masses and the radii 
decrease with the increase of the entropy for hadronic and quarkionic stars 
within the NJL model. For hybrid and MIT stars the behavior is not so well 
defined.

In the present work we have investigated one possible variation on the usual
electrically neutral pulsars: the inclusion of electric charge. 
We have observed that the behaviors shown in previous works with much
simpler EoS were also observed here. The influence of the temperature was also
investigated. We are now in a position to calculate the energy
released from the conversion of a metastable star (hadronic or hybrid) to a 
stable star (hybrid or quarkionic) under the influence of the electric charge. 
A more detailed and complete investigation, with a smaller bag parameter 
in the MIT bag model is under way.

\section*{Acknowledgments}
This work was partially supported by CNPq (Brazil). M.D.A. would like to thank
CNPq for a master's degree scholarship.

\end{document}